\documentclass[aps,prb,reprint,amssymb,twocolumn,superscriptaddress]{revtex4-1}

\usepackage{amsmath}
\usepackage{amsfonts}
\usepackage{amssymb}
\usepackage{graphicx}
\usepackage{moreverb}
\usepackage{url}
\usepackage{epsf}
\usepackage{color}
\usepackage{multirow}
\usepackage{subfigure}
\usepackage{morefloats}

\usepackage{cleveref}

\usepackage{threeparttable}

\usepackage{rotating}
\usepackage{cancel}
\usepackage{bm}
\begin{document}

\title{\emph{Ab initio} calculations of the optical absorption spectra of $\text{C}_{60}$-conjugated polymer hybrids}
\date{\today}

\author{Laura E. Ratcliff}
\email{laura.ratcliff@cea.fr}
\affiliation{Laboratoire de simulation atomistique (L\_Sim), SP2M, UMR-E CEA / UJF-Grenoble 1, INAC, Grenoble, F-38054, France}
\affiliation{Departments of Physics and Materials, Imperial College London, London, SW7 2AZ, United Kingdom}

\author{Peter D. Haynes}
\affiliation{Departments of Physics and Materials, Imperial College London, London, SW7 2AZ, United Kingdom}

\begin{abstract}
A recently developed linear-scaling density-functional theory (LS-DFT) formalism is used to calculate optical absorption spectra of hybrids of $\text{C}_{60}$ and the conjugated polymers poly(\emph{para}-phenylene) (PPP) and poly(\emph{para}-phenylene vinylene) (PPV).  The use of a LS formalism allows calculations on large systems with realistic proportions of $\text{C}_{60}$, which has been of interest for the use of such materials in photovoltaics.  Two different bonding structures are tested for the hybrid PPP and for both systems additional peaks are present in the absorption spectra below the original onset of absorption.  By identifying the eigenstates involved in the relevant transitions, a weighted density difference is formed, demonstrating the transfer of charge between the polymer chain and the $\text{C}_{60}$, in agreement with experiment.  For the hybrid PPV, no additional peaks are observed in the absorption spectrum.
\end{abstract}

\maketitle

\section{Introduction}

Since the discovery in 1976 that conjugated polymers could be made conducting~\cite{ppp_heeger,chiang1,shirakawa1}, they have been the subject of much research, both for reasons of fundamental theoretical interest and due to the promise of a wide range of exciting applications.  They are highly attractive candidates for use in applications ranging from light-emitting diodes (LEDs) to photovoltaic cells~\cite{ppp_burr,ppp_yu,facchetti_review,friend2001conjugated,ppp_moliton}.  This is in part due to the ability to influence their properties via doping and modification of side chains, enabling for example the creation of materials with a wide range of energy gaps resulting in materials ranging from metallic to insulating~\cite{chiang1,ppp_sac}.  Furthermore, they benefit from useful mechanical properties such as the ability to be flexible, and from desirable processing methods, particularly the ability to process from solution, which also has the important effect of reducing the final cost of the materials~\cite{ppp_sac,ppp_yu,ppp_heeger}.  As a result, a great deal of effort has been extended into developing and studying such polymers with the hope of replacing inorganic semiconductors in some existing applications as well as creating new ones.  

\begin{figure}
\begin{center}
\subfigure[poly(\emph{para}-phenylene)]{\label{fig:ppp_struc}\includegraphics[width=0.288\textwidth]{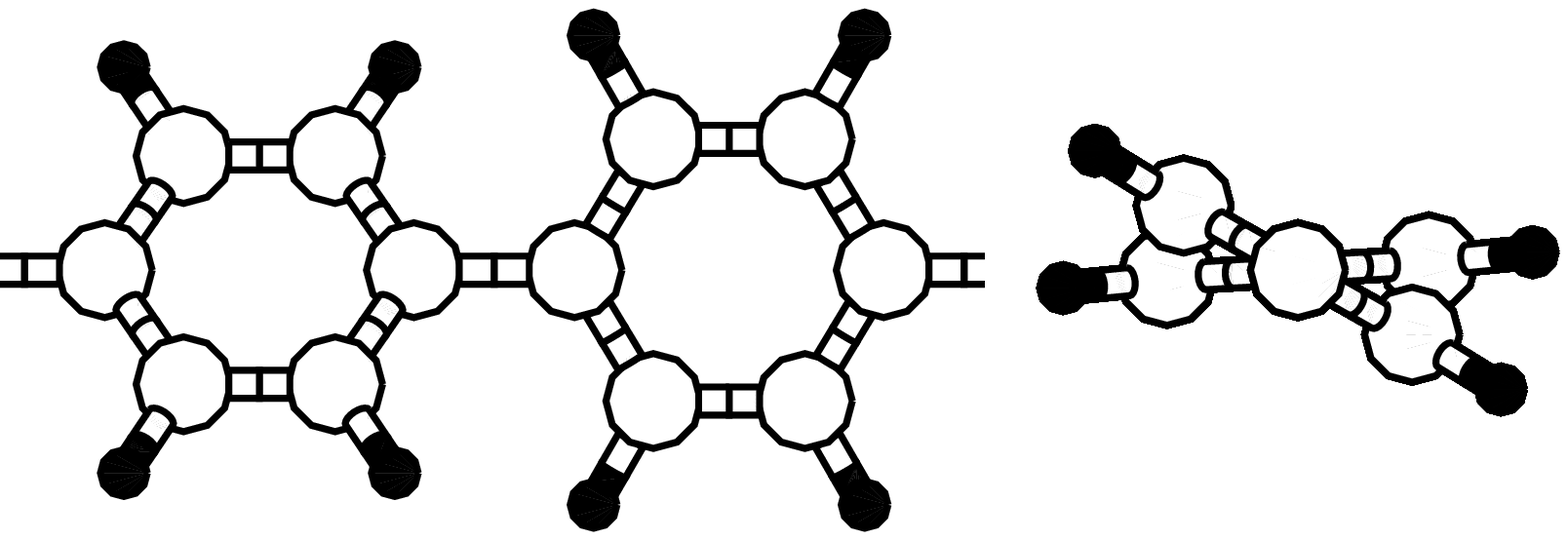}}
\subfigure[poly(\emph{para}-phenylene vinylene)]{\label{fig:ppv_struc}\includegraphics[width=0.27\textwidth]{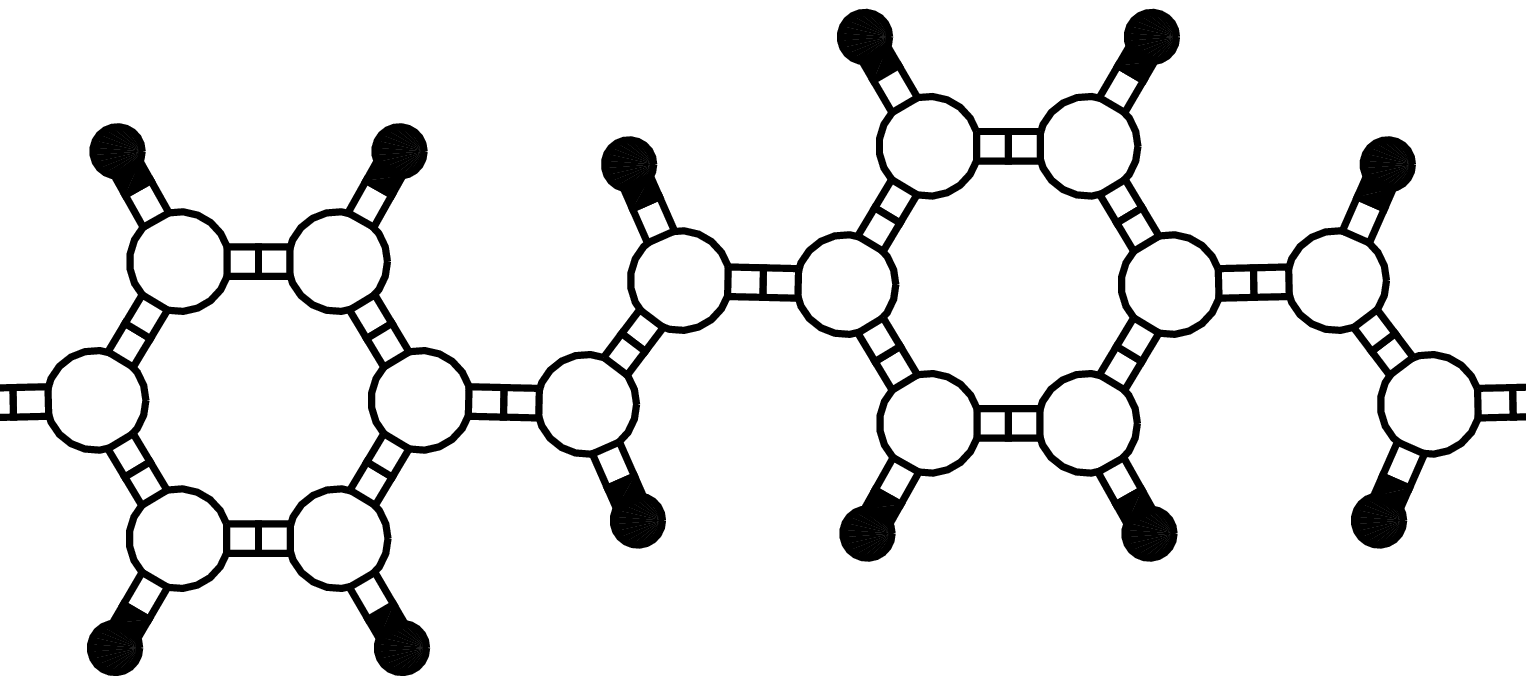}}
\caption{Schematics showing the structure of segments of PPP and PPV, with C atoms shown in white and H atoms in black.}
\label{fig:ppp_ppv_strucs}
\end{center}
\end{figure}

For the purposes of this work, we are particularly interested in the application to photovoltaic cells, where polymer based devices could have a number of advantages over the standard silicon based solar cell, particularly in terms of cost.  Indeed, conjugated polymers such as poly(\emph{para}-phenylene) (PPP) and poly(\emph{para}-phenylene vinylene) (PPV) (whose structures are depicted in Fig.~\ref{fig:ppp_ppv_strucs}) have been suggested as possible candidates for use as a donor when in combination with an acceptor such as the fullerene $\text{C}_{60}$~\cite{ppp_yu,ppp_moliton}.  For this use, certain criteria must be met, including the occurrence of charge transfer, which has been demonstrated in fullerene-polymer mixtures~\cite{ppp_sac,ppp_wang,smilowitz1993photoexcitation,ppp_lane}.  Indeed, $\text{C}_{60}$-PPV photovoltaic cells have been realized in a layered structure~\cite{halls1996exciton,halls1996exciton2}.

A useful way of studying these systems and comparing the behaviour of different conjugated polymers is through computational simulations, in particular first-principles calculations which allow the calculation of optical absorption spectra, among other useful properties.  One very popular methodology is density-functional theory (DFT)~\cite{hohenberg42,kohn43}, which has been applied extensively to a large number of systems and shown to give remarkably accurate results for ground state properties, and can also be used as a good first approximation to excited state properties such as absorption spectra.  The accuracy of these spectra will of course be limited by the approximations inherent in practical implementations of DFT, and arising from the fact that the Kohn-Sham eigenvalues have no rigorous correspondence with the true quasi-particle energies.  Indeed, one can only achieve good agreement with experiment by applying the ``scissor operator'', where the conduction states are rigidly shifted upwards in energy to correct for the underestimation of the band gap by DFT.  However, previous calculations on PPP have shown good agreement with the experimental absorption spectra using the local density approximation (LDA)~\cite{ppp_struc} and so we expect the use of DFT to be a reasonable approximation for these systems.  Furthermore, the systems of interest here involve many atoms, even when limiting ourselves to e.g.\ a single chain combined with a single molecule of $\text{C}_{60}$, so that the cost of more accurate methods such as the \emph{GW} approximation~\cite{hedin64,godby65,Sole97} is prohibitively expensive for calculations on systems of this size.  Indeed, even conventional DFT codes can become too expensive due to the cubic scaling bottleneck encountered for systems over a given size, and so it becomes useful to have a linear-scaling (LS) formulation of DFT, which takes advantage of the principle of nearsightedness and therefore allows one to access truly large systems.  

To this end we have employed a recently developed methodology for calculating first the Kohn-Sham conduction states and then the imaginary component of the dielectric function in a LS-DFT framework~\cite{my_long_paper,my_thesis}.  This new methodology combines the optimization of a second set of localized orbitals to describe the conductions states with the use of Fermi's golden rule and has been implemented in the LS-DFT code \textsc{onetep}~\cite{onetep1,onetep2,onetep3,onetep_forces}.  We have applied this methodology to the study of $\text{C}_{60}$ hybrids of PPP and PPV.

In this paper we first outline the basic theory behind the \textsc{onetep} code, including the method used to calculate the Kohn-Sham conduction states.  We then describe the method applied to the calculation of optical absorption spectra.  Finally we will present results for the absorption spectra of the hybrid PPP and PPV, before summarizing our conclusions.

\section{Methodology}

\textsc{onetep} is a linear-scaling DFT code which takes advantage of the principle of near-sightedness, and in particular the exponential decay of the density matrix (DM) in systems with a band gap~\cite{ismail46,kohn45,he81}.  This is achieved using a DM formulation, within a strictly localized representation of non-orthogonal generalized Wannier functions (NGWFs)~\cite{skylaris_ngwfs}.  The NGWFs are themselves represented in terms of a periodic cardinal sine (psinc) basis set~\cite{mostofi_prec}, which allows them to be optimized during the calculation.  The DM is written in separable form~\cite{mcweeny41,hernandez_kernel} as:
\begin{equation}
\langle\mathbf{r}|\hat{\rho}|\mathbf{r'}\rangle=\sum_{\alpha\beta}\phi_{\alpha}\left(\mathbf{r}\right){K^{\alpha\beta}}\phi_{\beta}^*\left(\mathbf{r'}\right) ,
\end{equation}
where $K^{\alpha\beta}$ is the so-called density kernel and $\{\phi_{\alpha}\left(\mathbf{r}\right)\}$ are the NGWFs.  An energy cut-off is applied to the psinc basis set in an analagous manner to plane-wave basis sets~\cite{onetep_dos_cond}, and thus \textsc{onetep} combines the advantage of a systematically improveable plane-wave basis set with the benefits of a localized basis set.  In a ground state energy calculation, both the DM and NGWFs are optimized to convergence, with the density and potential updated until self-consistency is achieved in the usual manner.  

During the calculation the eigenstates are not directly referenced, however they can be recovered via a single diagonalization at the end of the calculation.  This is an O$(N^3)$ process, however the prefactor is small as we optimize a minimal set of NGWFs that is much smaller than the number of psincs, so that the cost is negligible compared to the overall simulation time.  This enables for example the calculation of densities of states (DOS) and band structures.  However, as the NGWFs are optimized to minimize the total ground state energy, they are not expected to accurately represent the Kohn-Sham conduction states, and in some cases certain conduction states are found to be completely absent~\cite{onetep_dos_cond,my_long_paper,my_thesis}.  In order to be able to calculate optical absorption spectra one therefore needs a method of optimizing a new set of NGWFs to accurately represent the unoccupied states.  This is achieved by defining a projected Hamiltonian:
\begin{equation}
\hat{H}-\hat{\rho}\left(\hat{H}-\sigma\right)\hat{\rho},
\end{equation}
where $\hat{\rho}$ is the density operator and $\sigma$ is an energy shift.  The lowest eigenstates of this projected Hamiltonian are the lowest conduction states of interest and so it can be used in a similar procedure to ground state \textsc{onetep} calculations, optimizing a second set of NGWFs for a fixed number of conduction states, omitting the self-consistency process and instead re-using the ground state density and potential.  Following the conduction NGWF optimization, the valence and conduction NGWFs are combined to form a joint valence-conduction basis set, which is capable of representing both the occupied and unoccupied Kohn-Sham states.  Further details have been given elsewhere~\cite{my_long_paper,my_thesis}, including a demonstration of the linear-scaling behaviour of the method.

Once one has a process for calculating the Kohn-Sham conduction states as well as valence states, Fermi's golden rule can be applied to the calculation of optical absorption spectra.  The methodology~\cite{pickard72} used here is the same as that applied in the cubic-scaling plane-wave pseudopotential (PWPP) code \textsc{castep}~\cite{castep44}.  Using the dipole approximation, whereby only first order terms are included, one can write the imaginary component of the dielectric function as:
\begin{equation} \label{eq:imag_diel}
\varepsilon_2\left(\omega\right)=\frac{2e^2\pi}{\Omega\varepsilon_0}\sum_{\mathbf{k},v,c}\left|\langle\psi_{\mathbf{k}}^{c}|\mathbf{\hat{q}}\cdot\mathbf{r}|\psi_{\mathbf{k}}^{v}\rangle\right|^2\delta\left(E_{\mathbf{k}}^{c}-E_{\mathbf{k}}^{v}-\hbar\omega\right) ,
\end{equation}
where $v$ and $c$ denote valence and conduction bands respectively, $|\psi_{\mathbf{k}}^{n}\rangle$ is the $n$th eigenstate at a given $\mathbf{k}$-point with a corresponding energy $E_{\mathbf{k}}^n$, $\Omega$ is the cell volume, $\mathbf{\hat{q}}$ is the direction of polarization of the photon and $\hbar\omega$ its energy.  This expression can be directly applied for molecular systems but for extended systems one must use the momentum rather than position operator, as the position operator is ill-defined for periodic boundary conditions.  The correct relation between the two is~\cite{read_needs}:
\begin{equation}
\langle\psi_f|\mathbf{r}|\psi_i\rangle = \frac{1}{\text{i}\omega m}\langle\psi_f|\mathbf{p}|\psi_i\rangle + \frac{1}{\hbar\omega}\langle\psi_f|\left[\hat{V}_{\text{nl}},\mathbf{r}\right]|\psi_i\rangle ,
\end{equation}
where it is particularly important to include the second term, which involves the commutator between the non-local pseudopotential and the position operator.  In this manner one can calculate the imaginary component of the dielectric function and if desired the real component, using the appropriate Kramers-Kronig relation.  In practice, the joint DOS term is smeared using a Gaussian function, and the scissor operator is often applied to facilitate comparison with experiment.  

One could envisage constructing polymer/$\text{C}_{60}$ devices in a variety of ways with either covalent or non-covalent bonds connecting the polymer to the $\text{C}_{60}$, or by creating thin film devices.  Indeed there are a wide variety of structures which have already been realized for $\text{C}_{60}$-polymer hybrids~\cite{ppp_badamshina}.  We have chosen simple structures for hybrids of $\text{C}_{60}$ with PPP and PPV with concentrations of approximately 25\% by weight, in agreement with that used by Lane \emph{et al}.~\cite{ppp_lane} in their experiment on $\text{C}_{60}$-doped PPP.   We have calculated the optical absorption spectra for these hybrid structures, and compared the results to those of isolated chains of PPP and PPV and an isolated molecule of $\text{C}_{60}$. 

Before generating structures for the hybrid polymers, individual calculations were performed for the two polymers and the fullerene, for validation.  All calculations have been performed using the LDA exchange-correlation functional, with $\Gamma$ point sampling only, norm-conserving pseudopotentials and periodic boundary conditions.  The total energies were converged with respect to the psinc cut-off energy and NGWF radii -- for all calculations the psinc grid spacing was set to be equivalent to a kinetic energy cut-off of 1115~eV~\cite{onetep_dos_cond}, valence NGWF radii of 5.29~\AA\ and conduction NGWF radii of 7.41~\AA\ were used for all atoms, with four valence and conduction NGWFs per C atom and one valence and four conduction NGWFs per H atom.  No truncation was applied to the density kernel.

\section{Results and Discussion}

\begin{figure}
    \begin{center}
    \includegraphics[scale=0.55, angle=270]{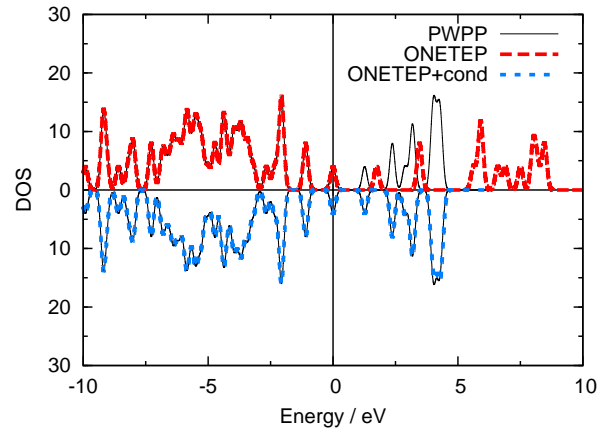}
  \caption{(color online) Density of states of PPV calculated with both \textsc{onetep} and a PWPP code.  In the upper panel, the PWPP result is plotted with the \textsc{onetep} result in the valence NGWF basis only, in the lower panel the PWPP result is plotted with the \textsc{onetep} result in the joint valence-conduction NGWF basis.  For each result, a Gaussian smearing width of 0.1~eV is used and only fifteen conduction states are included, which corresponds to the number which were optimized for the \textsc{onetep} conduction calculation.  Reproduced with permission from Ref.~\cite{my_thesis}.}
  \label{fig:ppv_dos_cond}
  \end{center}
\end{figure}

\begin{figure}
    \begin{center}
    \includegraphics[scale=0.55, angle=270]{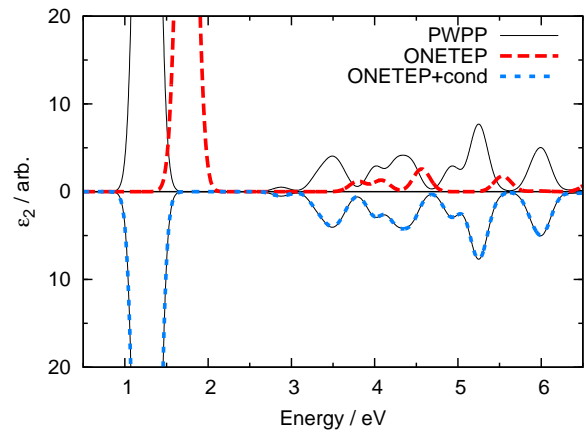}
  \caption{(color online) The imaginary component of the dielectric function of PPV calculated using the PWPP method and \textsc{onetep} both with (lower panel) and without (upper panel) a conduction calculation.  A Gaussian smearing width of 0.1~eV is used.  Reproduced with permission from Ref.~\cite{my_thesis}.}
  \label{fig:ppv_spec_cond}
  \end{center}
\end{figure}

In order to test the accuracy of the method compared to traditional plane-wave DFT, we have calculated the density of states and the imaginary component of the dielectric function for a small unit cell of PPV containing 56 atoms, using \textsc{onetep} both with and without the optimization of a set of conduction NGWFs.  These results are compared to PWPP results calculated using \textsc{castep} in Figs.~\ref{fig:ppv_dos_cond} and~\ref{fig:ppv_spec_cond} respectively.  The same pseudopotentials were used in both codes and the kinetic energy cut-off of the psinc and plane-wave basis sets were chosen to be approximately equivalent.  As can be clearly seen from these results, the optimization of a set of conduction NGWFs is vital for good agreement with traditional DFT results.  Further examples demonstrating the accuracy of the method are given elsewhere, together with a discussion on some of the limitations of the method~\cite{my_long_paper,my_thesis}.

\begin{figure}
\begin{center}
\subfigure[Structure A]{\label{fig:pppc60_struc}\includegraphics[width=0.4\textwidth]{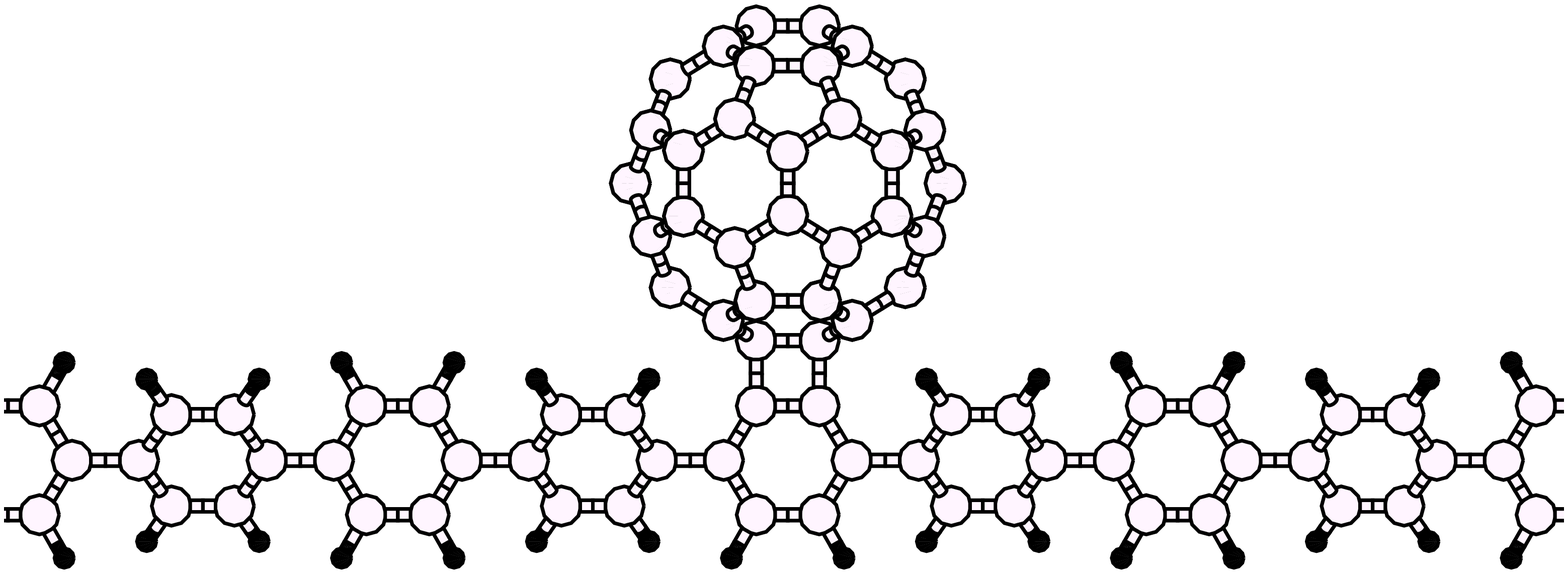}}
\subfigure[Structure B]{\label{fig:pppc60_struc2}\includegraphics[width=0.4\textwidth]{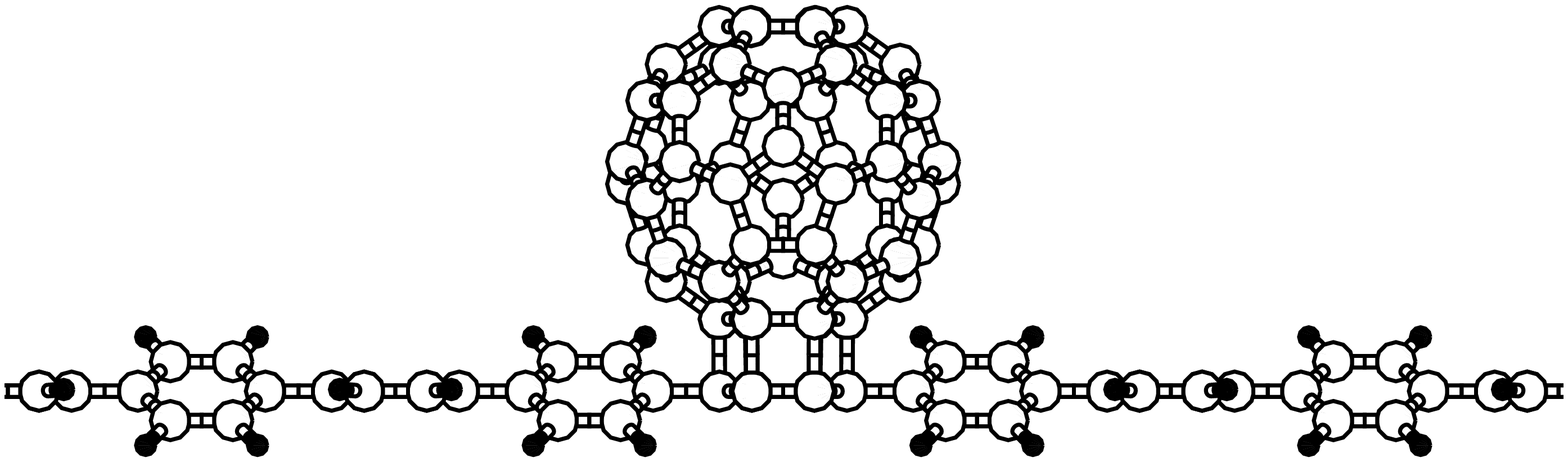}}
\subfigure[Structure B relaxed]{\label{fig:pppc60_struc2b}\includegraphics[width=0.4\textwidth]{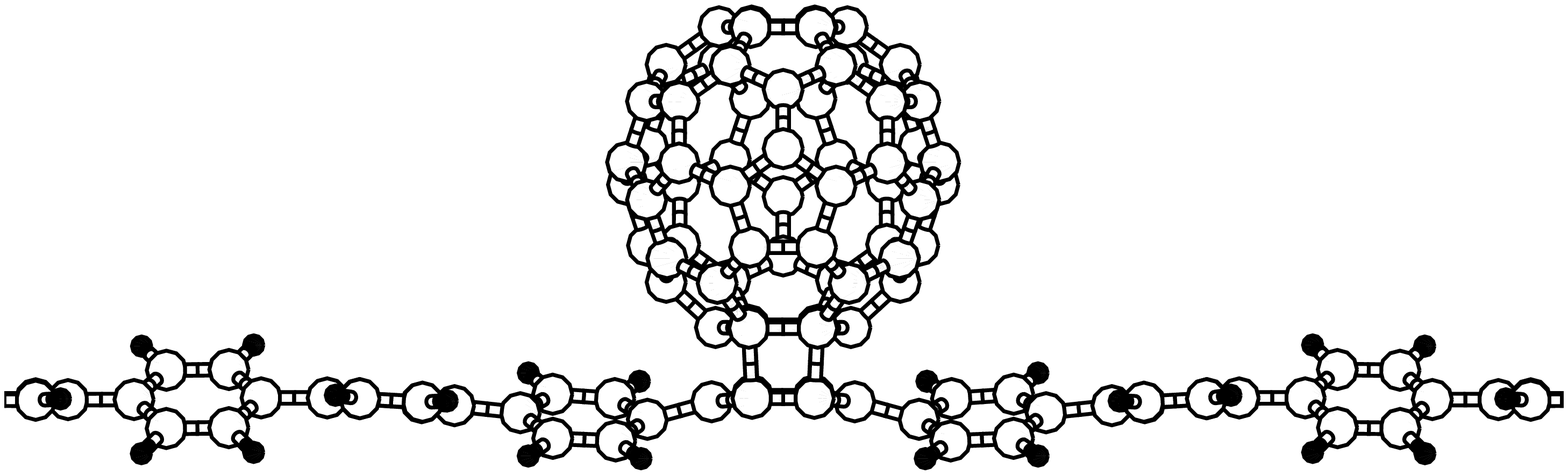}}
\caption{Schematics showing segments of the two different structures of $\text{C}_{60}$-PPP, including both the unrelaxed and relaxed forms of structure B.}
\label{fig:pppc60_strucs}
\end{center}
\end{figure}

Two different bonding structures were used for $\text{C}_{60}$-PPP, which are depicted in Figs.~\ref{fig:pppc60_struc} and~\ref{fig:pppc60_struc2}.  The $\text{C}_{60}$-PPV was constructed using the same attachment mechanism as structure A.  Unit cells of 676 (672) atoms were constructed in periodic boundary conditions for the hybrid polymers in structure A (B), with two molecules of $\text{C}_{60}$ evenly spaced across the polymers.  Ground state and conduction calculations were then performed in \textsc{onetep}, followed by the calculation of the imaginary component of the dielectric function.  The calculations on pure PPP and PPV were calculated in unit cells containing 280 and 560 atoms respectively, equivalent to samplings of 14 and 20 $k$-points with respect to the primitive unit cell, such that the spectra were converged.

\begin{figure*}
    \begin{center}
    \includegraphics[scale=0.55,angle=270]{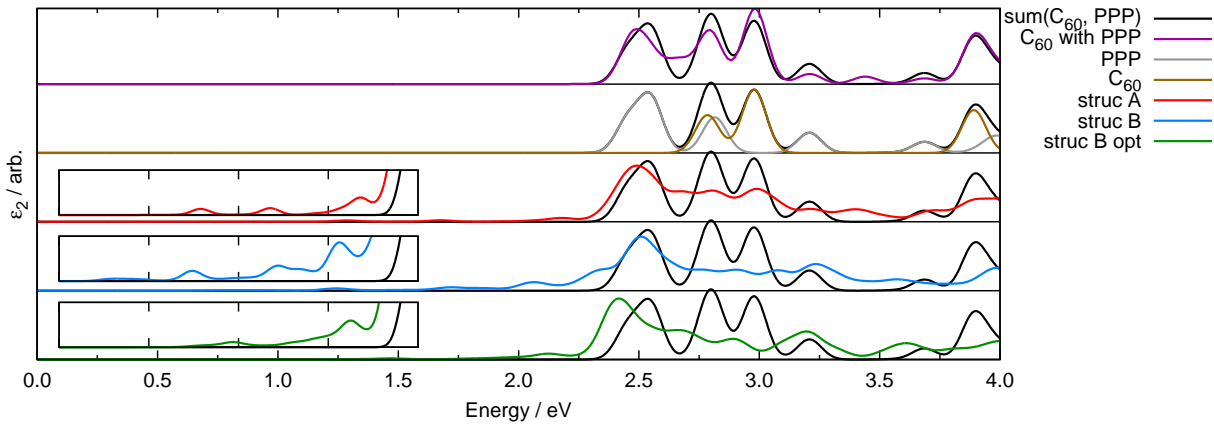}
  \caption{(color online) The imaginary component of the dielectric function of $\text{C}_{60}$-PPP in structures A and B, with both the unoptimized and optimized structure of B, compared to the sum of the spectra of pure PPP and $\text{C}_{60}$.  The spectra of pure PPP and $\text{C}_{60}$ are also compared with the total, as is that of $\text{C}_{60}$ and PPP in the same cell separated by a large distance (denoted $\text{C}_{60}$ with PPP in the key).  The three insets show a closer view of the spectra between 0.5 and 2.5~eV, highlighting the additional peaks.  The results are plotted with a Gaussian smearing width of 0.05~eV.}
  \label{fig:c60ppp_spec}
  \end{center}
\end{figure*}

The structures of both PPP and PPV were optimized using \textsc{castep} and the structure of $\text{C}_{60}$ was relaxed in the same manner.  To reduce the computational cost, the hybrid structures were initially not optimized.  The forces on the structures were calculated in \textsc{onetep}~\cite{onetep_forces}; the root mean squared (rms) forces for structures A and B of the hybrid PPP were 0.4 and 0.5~eV/\AA\ respectively and the maximum forces were and 4.1 and 4.7~eV/\AA\ respectively.  For the hybrid PPV they were 0.4 and 5.1~eV/\AA.  For the separated structures the forces were less than 0.1~eV/\AA.  As these forces for the hybrid structures are rather high, a test was done for structure B of $\text{C}_{60}$-PPP, by optimising the structure using \textsc{onetep} to give a rms force of 0.02~eV, and a maximum force of 0.1~eV.  The changes in the structure are essentially localized to the section of the chain closest to the fullerene, with the polymer being bent away from the $\text{C}_{60}$, as shown in Fig.~\ref{fig:pppc60_struc2b}.  The resulting spectrum was then compared to that of the unoptimized structure, giving an estimate of the accuracy of the results for the other structures (see Fig.~\ref{fig:c60ppp_spec}).

\begin{figure}
    \begin{center}
    \includegraphics[scale=0.55,angle=270]{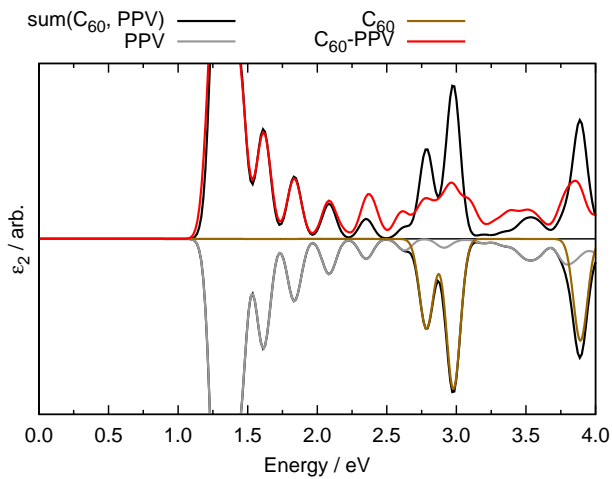}
  \caption{(color online) The imaginary component of the dielectric function of the $\text{C}_{60}$-PPV hybrid, compared to the sum of the spectra of pure PPV and $\text{C}_{60}$.  The spectra of pure PPV and $\text{C}_{60}$ are also compared with the hybrid spectrum in the lower half of the plot.  The results are plotted with a Gaussian smearing width of 0.05~eV.}
  \label{fig:c60ppv_spec}
  \end{center}
\end{figure}

Results for the different hybrid PPP structures are plotted in Fig.~\ref{fig:c60ppp_spec}, with the spectra of the PPP and $\text{C}_{60}$ shown for comparison.  The sum of the separate polymer and $\text{C}_{60}$ results is also plotted.  The results for both structures of $\text{C}_{60}$-PPP show a number of small additional peaks below the original onset of absorption of PPP (approximately 2.4~eV), which are not present in either PPP or $\text{C}_{60}$ on their own.  These peaks are stronger for structure B than structure A.  The relaxation of structure B has had a noticeable impact on the spectra, with changes both in position and relative height of the observed peaks.  In particular, those peaks which are below the original onset of absorption have significantly decreased in height, which is likely due to the bending of the polymer away from the fullerene.  However, the overall picture of additional peaks below this onset is still clearly preserved.  

The imaginary component of the dielectric function has also been calculated for $\text{C}_{60}$ and PPP in the same unit cell with the same cell dimensions as for the hybrid structures, but separated by a large distance (greater than 14~\AA) so that they are not chemically bonded, the result for which is also shown in Fig.~\ref{fig:c60ppp_spec}.  In this case, the relative heights of the peaks have been affected, and an additional peak is present at around 3.5~eV, however no additional peaks are present below the original onset of absorption.  This confirms that for those peaks below 2.4~eV to occur, the $\text{C}_{60}$ and PPP must be in close proximity.

\begin{figure*}
    \begin{center}
    \subfigure[HOMO-1 structure B]{\label{fig:homo_plot}\includegraphics[width=0.85\textwidth]{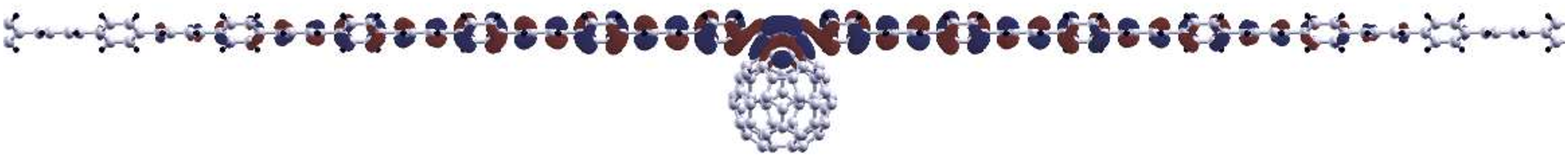}}
    \subfigure[LUMO+12 structure B]{\label{fig:lumo_plot}\includegraphics[width=0.85\textwidth]{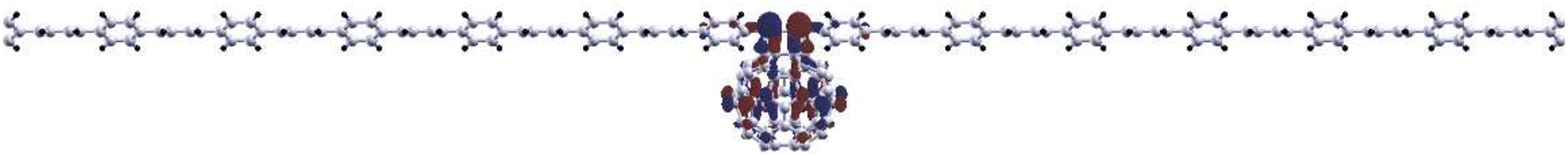}}
    \subfigure[Weighted density difference for structure B]{\label{fig:weighted_plot}\includegraphics[width=0.85\textwidth]{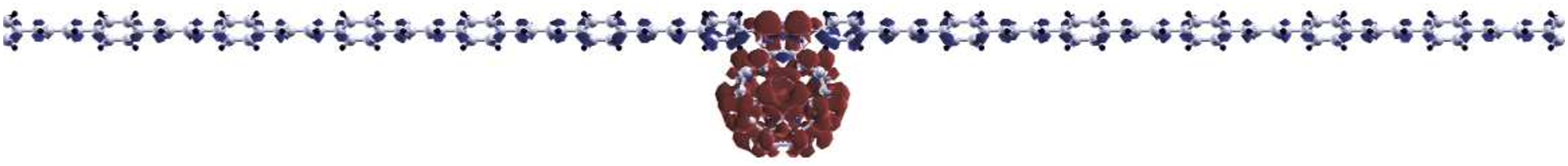}}
    \subfigure[HOMO-4 structure A]{\label{fig:struca_homo_plot}\includegraphics[width=0.85\textwidth]{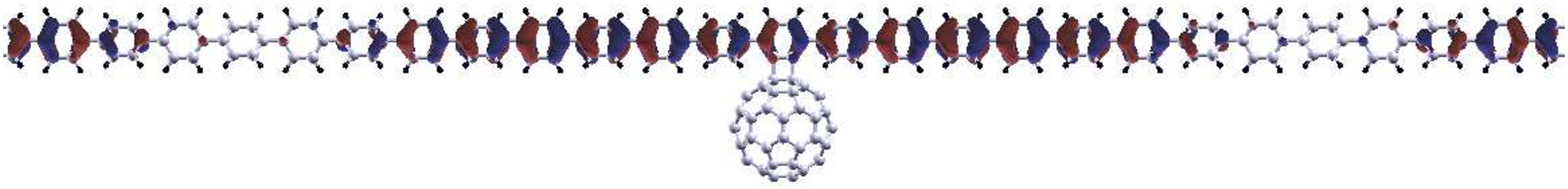}}
  \caption{(color online) Two different eigenstates for a segment of unrelaxed structure B of $\text{C}_{60}$-PPP, one occupied and one unoccupied, a weighted density difference for transitions below 2.4~eV for structure B and an eigenstate of structure A.  For the weighted density, red indicates an increase in electron density and blue indicates a decrease.  These figures were plotted using XCrySDen~\cite{Kokalj2003155}.}
  \end{center}
\end{figure*}

One of the advantages of computational spectroscopy is that it can be used to determine which transitions are responsible for a given peak.  In this way insight can be gained into the origins of the additional peaks in the hybrid PPP system.  For example, for the unoptimized structure B we can identify that there is a transition at 2.05~eV between the HOMO-1 (highest occupied molecular orbital) and the LUMO+12 (lowest occupied molecular orbital), for which the wavefunctions are plotted in Figs.~\ref{fig:homo_plot} and~\ref{fig:lumo_plot}.  We can see that the electron density is primarily located on the polymer for the HOMO-1, whereas for the LUMO+12 it is primarily located on the fullerene.  However, the three additional peaks below 2.4~eV result from a large number of transitions that are close together in energy.  Therefore, rather than attempting to visualize all of the individual transitions involved, as we have done for this example, it is more useful to generate an average density change, weighted by the strength of the transition involved.  To this end, we have used the following expression to calculate a weighted density difference $n_{\text{weighted}}$:
\begin{equation}
n_{\text{weighted}}=\frac{\sum_{v,c} \left|M_{cv}\right|^2 \left(\left|\psi_c\right|^2-\left|\psi_v\right|^2\right)}{\sum_{v,c}\left|M_{cv}\right|^2} ,
\end{equation}
where the sum is over all valence and conduction states which correspond to energy transitions in some given energy window and $M_{cv}$ is the optical matrix element for the transition between the corresponding valence and conduction state.  Fig.~\ref{fig:weighted_plot} shows the resulting density for transitions below 2.4~eV for the unoptimized structure B.  As can be seen, there is a significant increase in electron density on the $\text{C}_{60}$, whereas the electron density has decreased along the length of the chain.  This agrees with the expectation from experiment that charge transfer will occur between PPP and $\text{C}_{60}$ and suggests that the charge transfer process is stronger for structure B of the hybrid polymer.  This is possibly due to the fact that the $\text{C}_{60}$ is more strongly bonded to the chain in structure B, with six C-C bonds rather than two.  More importantly, we can see both from the bending of the chain during the relaxation of structure B and by visual inspection of the eigenstates close to the band gap that the pi-bonds on the phenyl ring adjacent to the fullerene have been disrupted, allowing for a strong mixing between states - an example of this can be seen in Fig.~\ref{fig:homo_plot}.  For structure A, on the other hand, the angle between the fullerene and the polymer is such that no similar disruption occurs, as shown for the HOMO-4 in Fig.~\ref{fig:struca_homo_plot}.  As a result, the additional peaks in the absorption spectrum for structure A are much less pronounced than those of structure B.

The spectrum for the hybrid $\text{C}_{60}$-PPV is shown in Fig.~\ref{fig:c60ppv_spec}, again plotted with those of PPV and $\text{C}_{60}$.  Unlike the PPP hybrids, the spectrum does not show any significant additional peaks, although some peaks have increased in intensity.  The lack of additional peaks in the spectrum for $\text{C}_{60}$-PPV could indicate that DFT alone is insufficient for capturing such behaviour in this particular system -- it has been previously demonstrated that the inclusion of excitonic effects has a strong impact on the absorption spectrum of PPV~\cite{Rohlfing2001101}.  However, an alternate study by Pedersen using density-functional-based tight-binding (DFTB) indicates that excitonic effects are important for both PPP and PPV~\cite{PhysRevB.69.075207} and so one also might not expect to see such peaks in $\text{C}_{60}$-PPP without their inclusion, which we can see is not the case.  Alternatively, it could be that the structure considered here is not ideal for the transfer of charge; in a further study by Pedersen~\cite{Pedersen20062488} charge transfer states are calculated in $\text{C}_{60}$-PPV with an arrangement similar to structure B, but with a finite chain length, a range of separation distances between the $\text{C}_{60}$ and the PPV greater than that used here and again using DFTB with the inclusion of excitonic effects.  Given that both the structure and formalism are different from those used in our calculation it is difficult to say which might have the bigger effect, however as the low energy peaks in structure A of $\text{C}_{60}$-PPP are already very small it seems more likely that it is the structure that is unfavourable, with too large an angle between the fullerene and the chain.  We can also see from the spectra that the states on $\text{C}_{60}$ are separated in energy from those of PPV, whereas with PPP there is a significant overlap in energy, particularly around 2.6~eV.  This should make it easy for these states to mix in the hybrid PPP structure to create a lower-lying state, which would not be the case for PPV.  The creation of such a state seems to be confirmed by the significant increase in the absorption spectra just below this energy seen for both PPP hybrids and for PPP and $\text{C}_{60}$ in the same simulation cell.  Finally, it could also be possible that there is some transfer of charge occurring between the PPV and the $\text{C}_{60}$, but it is not visible in the absorption spectrum either because it is too weak, or because it is masked by other peaks which are already present.

In practice, it is likely that chemically substituted PPP and PPV or other related conjugated polymers, such as poly[2-methoxy, 5-(2-ethylhexoxy)-1,4-phenylene vinylene] (MEH-PPV) or poly[2-methoxy-5-(3',7'-dimethyloctyloxy)-p-phenylene vinylene] (MDMO-PPV), might be used instead of pure PPP and PPV, due to the increased solubility that can be achieved in organic solvents, which is useful for easy fabrication of electronic devices.  Similarly, a lot of interest has been generated by fullerene derivatives such as [6,6]-phenyl-$\text{C}_{61}$ butyric acid methyl ester (PCBM) as an alternative to pure $\text{C}_{60}$ -- there have, for example, been a number of studies on MDMO-PPV in conjunction with PCBM, including the development of new techniques for experimental characterization~\cite{goris2005absorption,goris2006observation,PhysRevB.86.024201,faist2011competition}.  Furthermore, it would not necessarily be possible to easily manufacture the structures investigated in this work.  The above simplifications are justified as a first approximation in allowing us to model the absorption spectra of fullerene-conjugated polymer hybrids whilst keeping the computational costs low.  However, it would be interesting to extend this work in future to the consideration of other hybrids such as those mentioned above, which, despite the larger unit cells involved, will still be fully accessible within a LS-DFT framework.

\section{Conclusions}

In conclusion, we have investigated the effect on the optical absorption spectra of the attachment of molecules of $\text{C}_{60}$ to the two conjugated polymers PPP and PPV using a LS-DFT formalism, which has allowed us to simulate long chains with a realistic proportion of $\text{C}_{60}$ with respect to the polymer chain length.  For PPP, we considered two different bonding structures between the $\text{C}_{60}$ and the chain.  For both structures of the $\text{C}_{60}$-PPP hybrid additional peaks were present in the absorption spectra below the original onset of absorption.  By identifying the transitions between Kohn-Sham eigenstates which were responsible for these peaks and using the optical matrix elements to form a weighted density difference, we were able to demonstrate the transfer of electrons from the length of the chain to the $\text{C}_{60}$ molecule, agreeing with experimental observations of charge transfer.  For the PPV we did not find any additional peaks below the original onset of absorption, which is likely an indication that the bonding structure selected here is not optimal.  Further calculations could be done in future on different bonding structures of $\text{C}_{60}$-PPV such as structure B, in order to confirm this theory.

A number of approximations and simplifications were made, including the use of a purely DFT based formalism, which was justified based on the impracticality of going beyond DFT for such large systems.  In addition, relaxation of the hybrid structure was shown to affect the details of the absorption spectrum but not the overall picture and thus the use of unoptimized structures was also justified.  Furthermore, whilst the structures used were not necessarily those directly achievable in experiment or likely to be the precise systems of interest for photovoltaic systems given that interest has shifted from pure PPP and PPV to various derivatives, this work provides an ideal starting point for the study of large fullerene-polymer complexes.  Finally, this work could be easily extended in future to consider more complex polymers and fullerene-polymer combinations; indeed the linear-scaling formalism used here is ideal for the computational study of such materials and should prove to be useful in gaining further insights into systems which are of great interest for organic photovoltaics.

\begin{acknowledgments}
This work was supported by the UK Engineering and Physical Sciences Research Council (EPSRC). Calculations were performed on CX1 (Imperial College London High Performance Computing Service).  P.D.H. acknowledges support from the Royal Society in the form of a University Research Fellowship.
\end{acknowledgments}

\bibliographystyle{jolaura}
\bibliography{paper_pppc60}

\end{document}